В. Л. Бузько

Комунальний заклад «Навчально-виховне об'єднання №6 «Спеціалізована загальноосвітня школа І-ІІІ ступенів, центр естетичного виховання «Натхнення» Кіровоградської міської ради Кіровоградської області»

Ю. В. Єчкало

ДВНЗ «Криворізький національний університет»

# МОЖЛИВОСТІ ВИКОРИСТАННЯ QR-КОДІВ У НАВЧАННІ ФІЗИКИ

*Анотація*. У статті розглянуто можливості використання QR-кодів у навчанні фізики. Зазначено, що технології створення та розпізнавання QR-кодів можна віднести до елементів мобільного інформаційно-освітнього середовища. На основі узагальнення існуючих досліджень обговорюються переваги та недоліки використання QR-кодів, а також сфери застосування кодів у навчальному процесі. Наведено приклади використання QR-кодів у навчанні фізики (проведення фізичних квестів та веб-квестів; проведення ігор, вікторин, опитувань; створення віртуальної виставки; створення додатків до навчальних об'єктів; створення й дослідження комп'ютерних моделей фізичних явищ і процесів; організація самоперевірки). Встановлено, що мобільне навчання є доступним для учнів (студентів), а елементи мобільного інформаційно-освітнього середовища (зокрема, технології створення та розпізнавання QR-кодів) мають достатній потенціал у навчанні фізики.

*Ключові слова*: мобільне інформаційно-освітнє середовище, технології створення та розпізнавання QR-кодів, навчання фізики.

Процеси реформування, модернізації та розвитку різних соціальних сфер діяльності і галузей економіки, що розгорнулися в нашій країні в останні десятиліття, зумовлюють необхідність відповідних змін у вітчизняній системі освіти, переходу на новий рівень вимог до якості підготовки випускників шкіл та вузів. Пріоритетом розвитку освіти є впровадження сучасних інформаційно-комунікаційних технологій, зокрема – технологій та засобів мобільного навчання, які забезпечують удосконалення навчально-виховного процесу, доступність та ефективність освіти, підготовку молодого покоління до життєдіяльності в інформаційному суспільстві [2; 10].

Мобільне навчання є новою освітньою парадигмою, на основі якої створюється нове навчальне середовище, де учні (студенти) можуть отримати доступ до навчальних матеріалів у будь-який час та в будь-якому місці, що робить сам процес навчання всеохоплюючим та мотивує до безперервної освіти та навчання протягом усього життя. До основних переваг мобільного навчання можна віднести: можливість навчатися будь-де та будь-коли; компактність мобільних пристроїв; безперервний доступ до навчальних матеріалів; підвищену інтерактивність навчання; зручність застосування послуг мобільного навчання; персоналізованість навчання. Унікальними властивостями мобільного навчання є: придатність до одночасної взаємодії як з одним учнем (студентом), так і з групою; можливість динамічного генерування навчального матеріалу в залежності від місцезнаходження учнів (студентів), контексту навчання та способу використання мобільних пристроїв; можливість виконання окремих дискретних у часі навчальних дій учнів (студентів) у будь-який час і в будь-якому місці [10]. Розробці та використанню складових мобільних середовищ навчання присвячено роботи М. А. Кислової, В. О. Куклєва, Н. В. Рашевської, С. О. Семерікова, К. І. Словак, Ю. В. Триуса.

До складових елементів мобільного інформаційно-освітнього середовища входять мобільні інформаційно-комунікаційні технології і засоби навчання, до яких можна віднести технології створення та розпізнавання QR-кодів (від англ. quick response – швидкий відгук). Найбільш актуальним і затребуваним є використання QR-кодів у формуванні інформаційної складової навчального середовища й під час впровадження в освіті підходу BYOD (Bring Your Own Device – принеси свій власний пристрій). Практично будь-який мобільний пристрій легко розпізнає і розшифровує інформацію, закодовану за допомогою QR-коду.

Для цього потрібно лише піднести камеру мобільного пристрою зі встановленим програмним продуктом до зображення коду. Програма розшифрує код, а потім запропонує виконати певну дію, передбачену вмістом коду.

QR-коди є мініатюрними носіями даних, які зберігають текстову інформацію обсягом приблизно в половину сторінки формату А4. Можна закодувати текст, гіперпосилання, візитівку, повідомлення тощо. Ці дані кодуються за допомогою спеціальних програм або сервісів у вигляді чорно-білих або кольорових квадратів. QR-код містить в собі також додаткові дані, які потрібні для правильного декодування інформації спеціальними програмами мобільних телефонів чи інших пристроїв [1].

Як і для будь-якої іншої інновації, існує низка переваг та недоліків використання QR-коду (табл. 1) [4].

Таблиця 1

**Переваги та недоліки QR-кодів**

| Переваги | Недоліки |
|---|---|
| 1) зберігання великих обсягів цифрової та текстової інформації на будь-якій мові; | 1) відносно висока вартість мобільного Інтернету; |
| 2) швидкість створення QR-коду за допомогою програмних засобів; | |
| 3) висока швидкість розпізнавання, причому друкарський розмір коду може бути дуже малим; | 2) низький рівень поінформованості про технології QR-кодування; |
| 4) можливість зчитування в будь-якому напрямку; | 3) технічні неполадки. |
| 5) для розміщення підходить практично будь-яка поверхня; | |
| 6) стійкість до пошкоджень (зчитування при ушкодженні коду до 30%). | |

В тій чи іншій формі учні (студенти) щодня працюють з інформацією, здійснюючи її пошук, обробку, накопичення, передачу. Використання сервісів для створення та розпізнавання QR-кодів може надати допомогу викладачам як в аудиторній, так і в позааудиторній діяльності, сприятиме привернянню уваги школярів та студентів, їх зацікавленості, дозволить підвищити мотивацію. Де-які можливості використання QR-кодів у навчальному процесі показані у табл. 2, складеній нами за [1; 6].

Таблиця 2

**QR-коди у навчальному процесі**

| Сфера застосування | Результат діяльності |
|---|---|
| Гіперпосилання на мультимедійні джерела та ресурси | При супроводі заняття презентацією можна забезпечити слухачів роздатковим матеріалом з QR-кодами для доступу до допоміжних додатків (гіперпосилання на мультимедійні джерела та ресурси: відео-, аудіо-додатки, сайти, рисунки, анімації, електронні навчальні видання, бібліотеки тощо). Можна розмістити QR-коди й на самих слайдах презентації. Замість введення URL в свої телефони учні (студенти) можуть відсканувати код, щоб отримати додаткову інформацію миттєво. |
| Проектна діяльність | Під час організації проектної діяльності можна створювати колекції посилань, інформаційні блоки, коментарі на сторінках сайтів підтримки проекту, плакатах. Учні (студенти) можуть створювати власні портфоліо або анотації на прочитані книги та навчально-методичну літературу за досліджуваною темою й розміщувати їх на сайті проекту у вигляді QR-кодів. |
| Опитування та тестування | QR-коди дозволять організовувати швидкі опитування і проводити тестування як в аудиторії, так і поза нею (web-сервіси ClassTools, Plickers, Mentimeler та ін.). Наприклад, на кожному білеті з контрольним завданням можна розмістити надрукований QR-код з правильними відповідями або підказкою з алгоритмом розв'язування задачі. |
| Ігрові форми діяльності | QR-коди можуть бути використані в ігрових квестах для пропонування ігрових завдань на одному або декількох етапах відповідних заходів, у навчальних кросвордах. |
| Обкладинки навчально-методичної літератури | QR-коди доцільно використовувати для розміщення на обкладинках навчально-методичної літератури довідкового матеріалу, відомостей про автора, видавництво або будь-якої додаткової інформації. |
| Інформаційні стенди | QR-коди доцільно використовувати для інформаційного насичення стандартних інформаційних стендів у навчальних аудиторіях, лабораторіях, рекреаціях, бібліотеках, музеях навчальних закладів; для розміщення розкладу занять, результатів навчального процесу тощо. |
| Додатки до навчальних об'єктів | QR-коди можна розміщувати на частинах механізмів, електричних схемах, анатомічних об'єктах. Наприклад, розміщені на періодичній системі елементів QR-коди можуть містити фізичні та хімічні властивості елементів; розміщені на лабо- |

| | раторному (демонстраційному) обладнанні QR-коди можуть мати гіперпосилання на віртуальну лабораторію або контрольні запитання до самостійного опрацювання. |
|---|---|
| Ідентифікація | Розміщення контактної інформації на візитній картці викладача, адміністрації навчального закладу, на бейджах учасників конференцій (семінарів); ідентифікація учнів (студентів) у віртуальному кабінеті бібліотеки або дистанційного курсу. |

Наведемо приклади використання QR-кодів у навчанні фізики.

**1. Проведення фізичних квестів та веб-квестів.**

Фізичний QR-квест – це гра за типом лінійного квесту, в якому групам необхідно в умовах обмеження часу пройти якомога більше станцій і відповісти на питання, зашифровані за допомогою QR-коду. Клас ділиться на групи. Кожна група отримує:

– планшет зі встановленими на ньому програмою-декодером і презентацією для фотоміток;

– карту із зазначеними на ній станціями;

– робочий лист для записування відповідей.

Для того, щоб групи не заважали одна одній при проходженні маршруту, перші станції, з яких починається квест, у всіх груп різні, а надалі групи проходять точки-станції по порядку. На кожній станції групі необхідно розшифрувати код, відповісти на запитання і занести відповідь в робочий лист, зробити фотомітку і помістити її на відповідний слайд презентації.

Час квесту обмежений. Система оцінювання: за кожну пройдену станцію – 1 бал (за умови наявності фотомітки); за правильну відповідь – від 1 до 2 балів; за неправильну відповідь – 0 балів; за відсутність відповіді (за умови проходження даної станції) – штраф 1 бал; за запізнення на фініш – штраф 1 бал за кожні 5 хвилин запізнення.

Станції можуть бути різними. Під час узагальнюючого уроку «Теплові явища» (8 клас) учням варто запропонувати такі станції: «Історична», «Метрична», «Технічна», «Літературна». На нашу думку, доцільно запропонувати якісні задачі з фізики (слід обирати короткі завдання, інакше можуть виникнути складності із зчитуванням QR-кодів).

У процесі вивчення розділу «Основи термодинаміки» (10 клас) доцільно розглянути питання терморегуляції у живій природі, зокрема терморегуляцію людини. З цією метою варто реалізувати веб-квест «Терморегуляція в живій природі» [5]. Під час виконання веб-квесту учні поглиблюють свої знання з біології та фізики. Запропонований веб-квест сприяє осмисленню особистого досвіду спілкування дитини з природою і людьми; розумінню свого місця в природі та соціумі; важливість раціонального розуміння світу. Виконуючи завдання зазначеного веб-квесту, учні закріплюють знання з теплових явищ та вміння застосовувати одержані знання для розв'язування практичних задач, розширюють кругозір, удосконалюють навички користування Інтернетом для пошуку та обробки інформації, розвивають вміння самостійно працювати, раціонально розподіляти свій час [3]. Веб-квест створено відповідно до навчальної програми курсу фізики 10 класу [8] та оптимізовано відповідно до навчальної програми курсу фізики 8 класу [9]. Під час виконання такого веб-квесту учні виконують гру «Дешифрувальник». Учням необхідно дешифрувати «фізичну веселку» (рис. 1). Зашифровані завдання пропонують фантастичні персонажі Лірик і Мудрик.

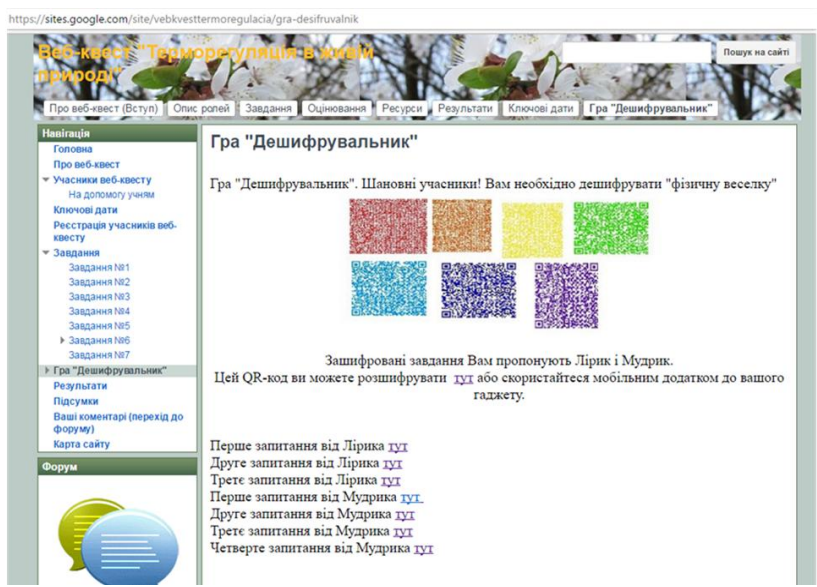

Рис. 1. Гра «Дешифрувальник»

Розглянемо приклад такого завдання (рис. 2). Учням пропонується: роз-шифрувати закодовані прислів'я; надіслати розшифровані записи за допомогою форми під QR-кодом; відповісти на запитання: «Про який вид теплообміну йде мова у прислів'ях?».

Відповіді на запропоновані завдання учні надсилають за допомогою Google Форм, які вбудовані в сторінку завдання.

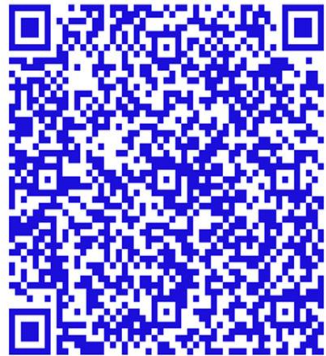

Рис. 2. Приклад закодованого запитання

**2. Проведення ігор, вікторин, опитувань.**

Під час проведення різноманітних ігор, вікторин, опитувань на одному з етапів заняття завдання може бути запропоноване у вигляді QR-коду. Прочитавши цей код, можна буде виконати завдання; роздрукувавши коди з необхідною інформацією, можна вклеїти їх безпосередньо в зошит учня (студента); контрольно-тестовий матеріал може бути виконаний у вигляді карток з різними варіантами завдань (рис. 3).

**3. Створення віртуальної виставки.**

Створення віртуальної виставки є доцільним, наприклад, під час вивчення теми «Теплові двигуни. Принцип дії теплових двигунів. ККД теплового двигуна» [9]. Вчителю потрібно підготувати зображення теплових машин на власному сайті, зашифрувати гіперпосилання на них у вигляді QR-кодів, роздрукувати коди з підписами та розвісити на стінах класу. Учні можуть у позаурочний час ознайомитися з експонатами виставки.

**4. Створення додатків до навчальних об'єктів.**

Учням (студентам) варто запропонувати розміщені на лабораторному обладнанні QR-коди, які містимуть гіперпосилання на віртуальну лабораторію або контрольні запитання до самостійного опрацювання. У кабінеті (лабораторії) фізики за допомогою QR-кодів можна надавати інформацію про призначення приладів, умов їх використання, історії створення, біографії їхніх творців.

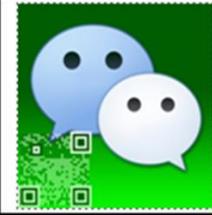

Рис. 3. Приклад картки для опитування за темою «Теплові явища»

**5. Створення й дослідження комп'ютерних моделей фізичних явищ і процесів.**

У ході виконання студентами практичних завдань доцільно, наприклад, запропонувати учням (студентам) побудувати графік функції розподілу Максвелла для кисню ($M$ = 0,032 кг/моль) при $T$ = 300 К [7] (рис. 4). Перед заняттям викладачу слід закодувати різні варіанти завдань для дослідження моделі.

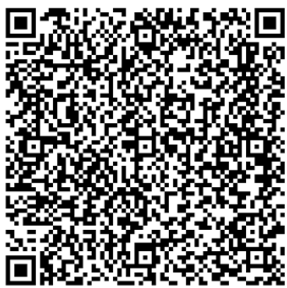

Рис. 4. Приклад завдання зі створення й дослідження
комп'ютерної моделі розподілу Максвелла

**6. Організація самоперевірки.**

QR-коди варто використовувати для організації занять з самоперевірки.

Учні (студенти) отримують список питань, правильні відповіді на які заздалегідь розміщені за різними Інтернет-адресами, представленим у вигляді QR-кодів.

Таким чином, на основі узагальнення існуючих досліджень, а також систематизації власних напрацювань, було визначено, що мобільне навчання є доступним для учнів (студентів), а елементи мобільного інформаційно-освітнього середовища (зокрема, технології створення та розпізнавання QR-кодів) мають достатній потенціал у навчанні фізики. Використання QR-кодів у процесі навчання фізики стимулює допитливість, інтерес учнів та студентів, активізує їх навчальну діяльність, водночас даючи змогу викладачам використовувати нові види навчальних пошуково-пізнавальних завдань узагальнюючої та систематизуючої спрямованості.

V. L. Buzko

Communal establishment «Educational Association №6 «Specialized School of I-III stages, aesthetic educational centre «Nathnennia» of Kirovohrad municipal council Kirovohrad region»

Yu. V. Yechkalo

SIHE «Kryvyi Rih National University»

THE POSSIBILITY OF USE OF QR-CODES IN TEACHING PHYSICS

*Abstract. In the article discusses the possibility of using of QR-codes in teaching physics. It is noted that the technology of recognition of QR-codes can be attributed to elements of mobile information and education environment. On the basis of summarizing existing research discusses the advantages and disadvantages of using QR-codes, and the application of codes in the learning process. Examples of the use of QR-codes in teaching physics (of physical quests and web quests, of games, quizzes, polls, creating a virtual exhibition, creating applications to educational facilities, the creation and study of computer models of physical phenomena and processes, organization Self-Test) are described. Found that the mobile learning available to pupils (students), and elements of the mobile information-educational environment (including technology development and recognition of QR-codes) have sufficient capacity in teaching physics.*

*Keywords: mobile information and education environment, technology of recognition of QR-codes, teaching physics.*

В. Л. Бузько

Учебно-воспитательное объединение № 6 «Специализированная общеобразовательная школа I-III ступеней, центр эстетического воспитания «Натхнення» Кировоградского городского совета Кировоградской области

Ю. В. Ечкало

ГВУЗ «Криворожский национальный университет»

ВОЗМОЖНОСТИ ИСПОЛЬЗОВАНИЯ QR-КОДОВ В ОБУЧЕНИИ ФИЗИКЕ

***Аннотация***. *В статье рассмотрены возможности использования QR-кодов в обучении физике. Отмечено, что технологии создания и распознавания QR-кодов можно отнести к элементам мобильной информационно-образовательной среды. На основе обобщения существующих исследований обсуждаются преимущества и недостатки использования QR-кодов, а также сферы применения кодов в учебном процессе. Приведены примеры использования QR-кодов в обучении физике (проведение физических квестов и веб-квестов; проведение игр, викторин, опросов; создание виртуальной выставки; создание приложений к учебным объектам; создание и исследование компьютерных моделей физических явлений и процессов; организация самопроверки). Установлено, что мобильное обучение является доступным для школьников (студентов), а элементы мобильного информационно-образовательного среды (в частности, технологии создания и распознавания QR-кодов) имеют достаточный потенциал в обучении физике.*

***Ключевые слова***: *мобильная информационно-образовательная среда, технологии создания и распознавания QR-кодов, обучение физике.*